\documentclass[%
 twocolumn,
 amsmath,amssymb,prc,aps,reprint,showpacs,floatfix]{revtex4-1}

\usepackage{bm}%
\usepackage{graphicx,array,multirow}
\usepackage{rotating}
\usepackage{hyperref}
\usepackage{url}
\usepackage{srcltx}
\expandafter \ifx \csname package@font\endcsname\relax\else
\expandafter \expandafter \expandafter
\usepackage
\expandafter \expandafter
\expandafter{\csname package@font\endcsname}%
\fi
\usepackage{soul}

\usepackage{epstopdf}

\newcommand{\el}{\ \\\nonumber}

\newcommand{\eg}{{\it e.~g. }}

\begin{document}

\title{On Recent Weak Single Pion Production Data}

\author{Jan T. Sobczyk and Jakub \.{Z}muda}

\affiliation{Institute of Theoretical Physics, University of Wroc\l aw, pl. M. Borna 9,
50-204, Wroc\l aw, Poland}

\date{\today}%
\pacs{13.15.+g,13.60.Le,24.10.Lx,25.30.Pt, CETUP* preprint number: CETUP2014-005}
\preprint{CETUP2014-005}
\begin{abstract}
MiniBooNE \cite{AguilarArevalo:2010bm} and MINERvA \cite{Eberly:2014mra} 
charge current $\pi^+$ production data in the $\Delta$ region 
are discussed. It is argued that despite the
differences in neutrino flux they measure the same dynamical mechanism of pion production
and should be strongly correlated. The correlation is clearly seen in the Monte Carlo simulations
done with NuWro generator but is missing in the data. Both normalization and the shape of the ratio of measured differential
cross sections in pion kinetic energy are different from the Monte Carlo results, in the case of normalization a discrepancy is  by a factor of $1.49$.
\end{abstract}

\maketitle

\section{Introduction}
\label{sec:spp}

There has been a lot of effort to understand better
single-pion-production (SPP) reactions in the neutrino-nucleon
and -nucleus scattering. These studies are motivated 
by the neutrino oscillation experiments 
demand to reduce systematic errors. In the few-GeV energy
region characteristic of experiments such as T2K \cite{Abe:2011ks},
NOvA \cite{Ayres:2004js}, LBNE \cite{Adams:2013qkq} and MicroBooNE \cite{Chen:2010zzt} the SPP channels
account for a large fraction of the cross section (at
1 GeV on an isoscalar target about 1/3 of the cross section).

In the 1 GeV energy region the dominant SPP mechanism is through $\Delta$ excitation.
There are several challenges in the theoretical
description of the SPP reaction in the $\Delta$ region. One comes from uncertainties
in the N-$\Delta$  transition matrix element, mainly due to the lack of precise information
on its axial part.
%The vector part was measured in photo- and electroproduction experiments, but precise information on its axial counterpart is missing.
In order to describe the SPP channels one also needs a significant nonresonant background contribution. 
Several theoretical models have been developed (see Refs.  \cite{Adler,*Fogli:1979cz,*Rein:1987cb,*Sato:2003rq,*Sato:2004he,*Hernandez:2007qq,*Mariano:2007zz}) 
to predict its shape and magnitude.
The differences between them introduce important model dependence in the N-$\Delta$ 
transition matrix element analysis and even in the description
of the $\Delta$ resonance. 
%This causes a confusion between physicists trying to reproduce single pion production cross sections.

In theoretical computations of SPP on atomic nuclei nuclear effects must be incorporated starting from 
the Fermi motion and Pauli blocking. It is very important to entail
the in-medium $\Delta$ self-energy. Its real part shifts the pole, whereas the
imaginary part corresponds to medium-modified SPP
and pionless $\Delta$ decay processes. The problem of charge current SPP on nuclei has
been addressed in Refs. \cite{Singh:1998ha,*Ahmad:2006cy,*SajjadAthar:2009rc} by assuming the $\Delta$ dominance
with many-body effects taken from Ref. \cite{Oset:1987re}. The
computations suggest a significant reduction of the pion
production cross section.

On top of all that, in the impulse approximation regime 
final state interactions (FSI) effects must be carefully evaluated, see e.g. Ref. \cite{Buss:2011mx}.
FSI include: pion rescattering, absorption, charge
exchange, and (for sufficiently
high energies) production of additional pions. The nuclear physics uncertainties are so
large that in most of the cases experimental groups do not try to
measure the characteristics of a neutrino-nucleon SPP process.
They publish instead the cross section results with all the nuclear
effects included with signal events defined by outgoing pions.

More precise SPP measurements on both nucleon and nucleus target
are necessary. The models of $\Delta$ excitation matrix
elements and nonresonant background are still validated
mainly on the old low-statistics bubble chamber experiments
performed at Argonne National Laboratory (ANL, \cite{Barish:1978pj,*Radecky:1981fn})
and Brookhaven National Laboratory (BNL, \cite{Kitagaki:1986ct,*Kitagaki:1990vs}). The nonresonant
background is more important in neutrino-neutron SPP channels where the cross sections are smaller
than for neutrino-proton SPP reaction and the statistical
uncertainties are larger \cite{Lalakulich:2010ss, Graczyk:2014dpa}. 

In view of those limitations it is important to explore the information 
from more recent neutrino-nucleus cross section measurements.
In the case of CC $1\pi^+$ reaction on the Carbon target
interesting studies were done by MiniBooNE \cite{AguilarArevalo:2010bm} and MINERvA \cite{Eberly:2014mra} experiments.
Both analyses focus on the $\Delta$ region. The main difference is in the neutrino energy. 
Typical MiniBooNE interacting neutrinos energies are smaller by a factor of $\sim 4-5$.

The main results of this paper are following. According to Monte Carlo simulations 
a strong correlation between the differential cross sections in pion kinetic energy in two experiments
is expected. The MINERvA cross section is expected to be larger by a factor of $\sim 2$. The shape of differential cross sections 
is anticipated to be very similar. This 
correlation is absent in the published data. The data/Monte Carlo discrepancy is seen in a particularly clear way when one compares 
the ratio of differential cross sections from two experiments with the Monte Carlo predictions. The experimental quantity is far from the anticipated
value of $\sim 2$. Also the shapes of predicted and measured ratio are different.

The paper is organized as follows:
In section \ref{sec:data} MiniBooNE and MINERvA SPP data are discussed and
re-binning of MniBooNE data according to the MINERvA bins is done.
Monte Carlo-data comparison is presented in section \ref{sec:nuwro} with the main
result shown in paragraph \ref{sec:ratio}.
In section \ref{sec:conclusions} we conclude our paper.

\section{MiniBooNE and MINERvA SPP data}
\label{sec:data}

The MiniBooNE measurement was done on the mineral oil target ($CH_2$). The neutrino flux
peaks at $\sim 700$~MeV
with a tail extended to $3$~GeV. The signal charged current
events are defined as $1\pi^+$ and no other mesons in the final state.

MINERvA measurement was done on the $CH$ target with larger energy flux peaked at 
$\sim 3$~GeV and a long high energy tail.
The signal charged current events contain exactly one charged pion, almost always $\pi^+$.
Due to the cut on invariant hadronic mass
$W<1.4$~GeV a contamination from the $1\pi^\pm 1\pi^0$ events is very small.

In both cases the signal includes a fraction of coherent $\pi^+$ production events.

Even if in the case of the MiniBooNE measurement typical neutrino energies are 
lower by a factor of $\sim 4-5$ compared to the MINERvA, in both experiments the dominant 
contribution comes from $\Delta(1232)$ excitation. In both cases the target consists mostly
from Carbon and we expect a lot of similarity in the measured cross sections.
According to NuWro Monte Carlo simulations the only significant difference in both measurements
comes from overall normalizations. Typical MiniBooNE $\nu_\mu$ energies are
closer to the pion production threshold energy. From the ANL and BNL experiments it is known that with a cut
$W<1.4$~GeV $\pi^+$ production cross sections at $\sim 700$~MeV and $\sim 4$~GeV differ
by a factor of $\sim 2$.

\begin{figure}[!htb]
\centering\includegraphics[width=\columnwidth]{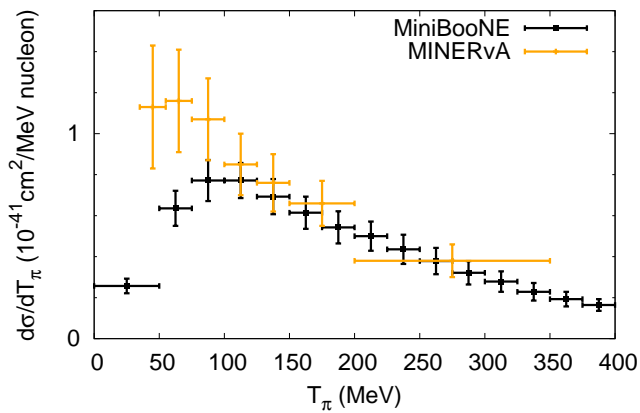}
\caption{(Color online) Differential cross section in pion kinetic energy. MiniBooNE data points are taken from Ref. \cite{AguilarArevalo:2010bm}
and MINERvA from Ref. \cite{Eberly:2014mra}.}\label{fig1}
\end{figure}

The cross section results from two experiments are shown 
in Fig. \ref{fig1}. The errors are given as a fractional uncertainty for each bin.
Both experiments have different binning. MiniBooNE binning is finer
than MINERvA, hence we will ``translate'' the MiniBooNE data to the MINERvA bins in order to
perform a direct comparison.

In Fig. \ref{fig1} we see, that in most of the cases MINERvA bins overlap with two MiniBooNE bins at most.
We use a linear interpolation of cross section and its error. The measured points
can be correlated, but there is no available information about the covariance matrix for considered data.
This procedure is justified if the new bin contains data from two bins and can not be applied to combine higher number of bins.

In the latter case we use the following method.
We assume that each data point represents a random variable with expected value equal 
to the central value (cross section in i-th bin, $E(X_i)=\sigma(E_i)$) and variance equal to the squared error
($Var(X_i)=(\Delta\sigma_i)^2$).
i-th MiniBooNE bin contributes to j-th MINERvA bin with a weight equal to the ratio 
of the bin's intersection $\alpha_{ij}$ with the MINERvA bin to the MINERvA bin width $W_j$:

\begin{eqnarray}
\label{eq:weight}
w_{i,j}= \frac{\alpha_{i,j}}{W_j}.
\end{eqnarray}
The expected value of the MiniBooNE cross section in the j-th MINERvA bin is:
\begin{eqnarray}
\label{eq:expect}
E(Y_j)=E(\sum_i w_{ij}X_i)=\sum_i w_{ij} E(X_i).
\end{eqnarray}
In the above equation $E(X_i)$ represents the measured MiniBooNE cross section.
The variance of the sum of $N$ random variables is:
\begin{eqnarray}\nonumber
Var(\sum_{i=1}^N w_{i,k} X_i)&=& 
\sum_{i=1}^N w_{i,k}^2 Var(X_i)+\\\label{eq:varcov} &+& 2\sum_{i> j}w_{i,k}w_{j,k}Cov(X_i,X_j).
\end{eqnarray}

Unfortunately, MiniBooNE experiment did not publish
the covariance matrix. The experimental errors are almost entirely systematic. 
%The biggest contribution comes from the neutrino
%flux uncertainty, which cannot be reduced by combining neighboring bins. 
%The second biggest error comes from the background estimation. 
The simplest assumption $Cov(X_i,X_j)= 0$
would reduce the error during the rebinning operation, since $w_{ij}\leq 1$. 
A reasonable assumption for these systematic errors
is that if one combines the neighboring bins the resulting error is a weighted 
average of contributing bin errors.
It is easy to show, that if one sets $Cov(X_i,X_j)=\sqrt{Var(X_i)Var(X_j)}$ the resulting 
error will be exactly a weighted average:
\begin{widetext}
\begin{eqnarray}\nonumber
Var(\sum_{i=1}^N w_{i,k} X_i)&=& \sum_{i=1}^N w_{i,k}^2 
Var(X_i)+2\sum_{i> j}w_{i,k}w_{j,k}\sqrt{Var(X_i)Var(X_j)}=\\\label{eq:varpen}&=&
\sum_{i=1}^N w_{i,k}^2 (\Delta\sigma_i)^2+
2\sum_{i> j}w_{i,k}w_{j,k}\Delta\sigma_i\Delta\sigma_j=
\left(\sum_{i=1}^Nw_{i,k} \Delta\sigma_i\right)^2.
\end{eqnarray}
\end{widetext}

\begin{figure}[!htb]
\centering\includegraphics[width=\columnwidth]{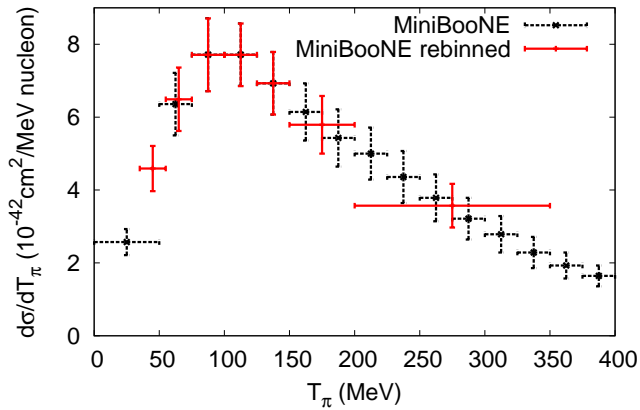}
\caption{(Color online) MiniBooNE differential cross section in pion kinetic energy taken from Ref. \cite{AguilarArevalo:2010bm} rebinned into MINERvA-sized bins.}\label{fig2}
\end{figure}

The new data binning for the MiniBooNE experiment according to Eqs. \ref{eq:expect} and \ref{eq:varpen} is shown in Fig. \ref{fig2}.
%We made an exception in our procedure for a few points with the lowest pion kinetic energy. The reason is that in this case
%one combines these bins from two neighboring MiniBooNE data points, with a very little overlap with one of them. 
%We replace Eqs. \ref{eq:expect} and \ref{eq:varpen} with simple linear interpolation.
%This procedure is justified if the new bin contains data from two bins and can not be applied to combine higher number of bins.
After re-binning the systematic errors have not been reduced, as expected. 

\section{NuWro Monte Carlo event generator}
\label{sec:nuwro}

NuWro is a versatile Monte Carlo simulation tool describing lepton-nucleon and lepton-nucleus 
interactions in the energy range from $\sim 100$~MeV to $1$~TeV. 
Its main functionalities and implemented physical models are presented
in Ref. \cite{Golan:2012wx}. Neutrino-nucleon interaction
modes are: quasi-elastic (or elastic for NC)
(QEL), resonant (RES) covering $W<1.6$~GeV and DIS defined by $W>1.6$~GeV. 
For the purpose of this study the 
anisotropy in the pion angular distribution in $\Delta$ decay events 
has been implemented using the density matrix elements measured by
ANL \cite{Barish:1978pj,*Radecky:1981fn} and BNL \cite{Kitagaki:1986ct,*Kitagaki:1990vs} experiments.

For the neutrino-nucleus scattering impulse approximation is assumed.
New reaction modes, absent in neutrino-nucleon scattering, are:
coherent pion production (COH) and
two-body current interactions on correlated nucleon-nucleon pairs (MEC).
In our simulation we used Valencia MEC model 
\cite{Nieves:2011pp} with the  momentum transfer
cut $|\vec{q}|<1.2$~GeV, as suggested in 
Ref. \cite{Gran:2013kda}. 
In MEC events final state nucleons are described using a model proposed
in Ref. \cite{Sobczyk:2012ms}.

Primary interaction is followed by hadron rescatterings (FSI) simulated by 
the custom made internuclear cascade model \cite{Golan:2012wx}.

In the simulations discussed in this paper
Carbon nucleus is treated within the relativistic Fermi Gas model. 
$\Delta$ in-medium self-energy effects are included in an approximate way using
the results from Ref. \cite{Sobczyk:2012zj}.

According to NuWro simulations, in the MiniBooNE and MINERvA experiments the pion production signal events origin from:
\begin{enumerate}
\item RES interactions, typically through the $\Delta$ excitation and decay, 
but also with some contribution from the nonresonant background. According to NuWro
RES accounts for $87.1\%$ and $84.7\%$ of the signal for the MiniBooNE and MINERvA experiment respectively.
There is a very important impact of FSI effects on the final state pions
production rate because many pions are absorbed or suffer from the charge exchange reaction
inside Carbon nucleus.

\item COH process, populating  $6.7\%$ (MiniBooNE) and $10.7\%$ (MINERvA) of the signal. 
NuWro uses the Rein-Sehgal coherent pion production model from Ref. \cite{Rein:1982pf} with lepton mass correction from Ref. \cite{Berger:2008xs}.
A comparison with the recent MINERvA coherent pion production measurement published in Ref. \cite{Higuera:2014azj}
suggest that NuWro may overestimate the experimental data. 

\item DIS interactions contributes only to the MiniBooNE signal at the level of $3.6\%$.
A typical scenario is that one out of two pions 
produced in the primary interaction is absorbed.

\item QEL and MEC interactions with pions produced due to the
nucleon rescattering reactions account for $2.7\%$ MiniBooNE and $4.6\%$ MINERvA signal events.

\end{enumerate}

\begin{figure}[!htb]
\centering\includegraphics[width=\columnwidth]{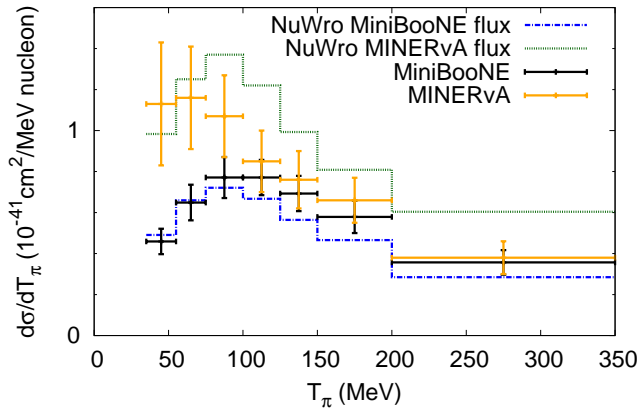}
\caption{(Color online) Differential cross section in pion kinetic energy. MiniBooNE and MINERvA data points are shown together with
NuWro predictions.}\label{fig3}
\end{figure}

Results of the NuWro simulations together with the experimental points are shown in  Fig. \ref{fig3}. 
The Monte Carlo predictions tend to overestimate 
the MINERvA data and underestimate the MiniBooNE data at the same time. 
In the case of MiniBooNE results similar problems were reported in the past
by many theoretical models (see \eg Refs. \cite{Lalakulich:2012cj} and \cite{Rodrigues:2014jfa}).
Another observation is that the MC simulation predicts a large difference between
the MiniBooNE and MINERvA cross sections in the whole pion kinetic energy range. 
On the other hand, in Fig. \ref{fig3} 
one can see, that for higher pion kinetic energies values reported by both experiments are very similar. Also, in the MC simulations 
the differential cross sections tend to peak at the same point 
in pion kinetic energy, near the threshold for 
$\Delta$ production, which in the pion FSI simulations leads to significant
pion absorption. However, both of the experimentally measured 
cross sections seem to reach their maximal values at different points. This is not pronounced very strongly
in Fig. \ref{fig3}, because the MINERvA errors are very large. 
We checked that introduction of the anisotropy for the pion angular distribution
does not change much NuWro results, giving an effect of at most  
10\% in a few kinetic energy bins but typically much smaller. The shapes of differential
cross sections changes a little but there is almost no structure to it save for the kinetic energy distribution.
We observe there a shift by $\sim$5\% of of the cross section towards higher kinetic energies in both MINERvA
and MiniBooNE distributions.

NuWro results look consistent with the GENIE \cite{Andreopoulos:2009rq} predictions for
$\frac{d\sigma}{dT_\mu}$, shown in Fig. 4 of Ref. \cite{Eberly:2014mra}. 
NuWro and GENIE use different physical models to describe the
SPP (GENIE relies on the resonant Rein-Sehgal model)
but both predict a strong correlation between results from two experiments.

%%%%%%%%%%%%%%%%%%%%%%%%%%%%%%%%%%%
\subsection{Ratio of MINERVA/MINIBOONE cross section in pion kinetic energy}
\label{sec:ratio}
%%%%%%%%%%%%%%%%%%%%%%%%%%%%%%%%%%

Correlations of both $\pi^+$ production measurements should be clearly seen in the ratios of measured 
differential cross sections in pion kinetic
energy relative to the neutrino flux from two experiments. Their shapes do not 
depend on the overall normalizations in two experiments.

In order to calculate the ratios of both measurements together with appropriate errors
we consequently treat the processed data points as random variables $X$ 
and $Y$ with known expected values
and variances. One has to compute $E(\frac{X}{Y})$ and $Var(\frac{X}{Y})$. 
For independent variables:
\begin{eqnarray}
\label{eq:expectprod}
E(X\cdot Z)&=&E(X)E(Z)\el
Var(X\cdot Z)&=&Var(X)Var(Z)+\\\label{eq:varprod}&+&E(X)^2Var(Z)+E(Z)^2Var(X)
\end{eqnarray}
and replacement $Z=\frac{1}{Y}$ must still be done. 

The assumption that two experiments are independent is rather
conservative because 
errors coming from neutrino interaction models are in both cases correlated.

The most difficult task is
to calculate $E(\frac{1}{Y})$ and $Var(\frac{1}{Y})$, 
because $E(\frac{1}{Y})\neq\frac{1}{E(Y)}$ unless the probability distribution function of $Y$ is given by 
the Dirac delta function, $P(Y)=\delta(Y-Y_0)$. We
must introduce some model-dependence, which fortunately will be shown to be negligible. 

We investigated several assumptions for the $P(Y)$:
\begin{itemize}
\item flat distribution
\item linear distribution
\item quadratic distribution
\item log normal distribution
\end{itemize}

The assumption is that $P(Y\leq 0)= 0$
and $P(Y)$ drops to 0 faster, than $Y^2$ as $Y$ approaches 0 since the cross section 
cannot be negative and we do not want the integral
to give indefinite values for $E(1/Y)$ and $E(1/Y^2)$.

We tested the model dependence of ratios using above probability 
distribution hypotheses by 
calculating both the expected ratio value as well as its error. We compared them also
to a ``naive'' approximation, in which $E\left(\frac{1}{Y}\right)\approx\frac{1}{E(Y)}$ 
and $Var\left(\frac{1}{Y}\right)\approx\frac{1}{Var(Y)}$.

We verified that the expectation values and variances coming from various
probability distribution hypotheses do not differ in any significant manner. 
The only exception is the ``naive'' approach,
leading to a few-percent effect on the expected value and enlargement of the variance.
From the above described models we chose the log-normal distribution as it allows 
any value of random variable along the positive real semiaxis. It has the following
functional form, expected value and variance:

\begin{eqnarray}\nonumber
P(Y)&=&\frac{1}{\sqrt{2\pi}bY}\exp\left[-\frac{(\ln(Y)-a)^2}{2b^2}\right]\Theta(Y)\el
E(Y)&=&\exp(b^2/2+a)\\\label{eq:lognormal}
Var(Y)&=&\exp(2b^2+2a)
\end{eqnarray}

From the above equation we get $E(\frac{1}{Y})=\exp(b^2/2-a)$ and 
$Var (\frac{1}{Y})= \exp(b^2-2a)\left[\exp(b^2)-1\right]$.

The procedure is to generate samples with NuWro generator for both experiments
and compare the resulting ratio of Monte Carlo cross sections to the experimental one.
In order to maintain statistically meaningful samples each dynamical channel contributing
to the MINERvA and MiniBooNE signals has been generated separately.

We tried to estimate the errors of both ratios as calculated by NuWro.
We distinguish systematic and statistical errors coming from the
implemented theoretical models. We run the simulations with a high event rate
in order to minimize statistical fluctuations. We obtained at least 8\;000 events in each bin
with typical value of the order of $10^4-10^5$ events/bin. The resulting impact of statistical errors
on the predicted ratio is negligible.

\begin{figure}[!htb]
\centering\includegraphics[width=\columnwidth]{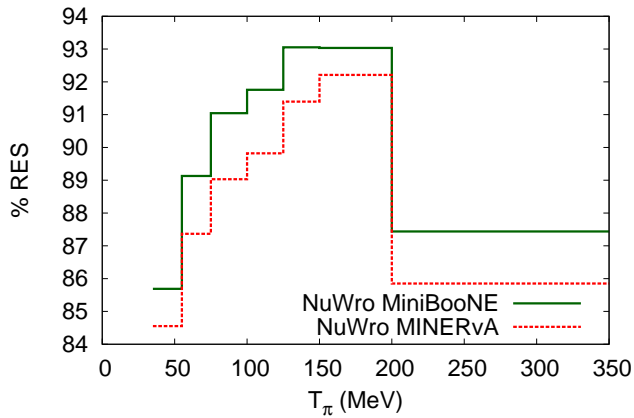}
\caption{(Color online) Contributions from RES channel to MiniBooNE and MINERvA differential cross sections in pion kinetic energy
as predicted by NuWro.}\label{fig4}
\end{figure}

\begin{figure}[!htb]
\centering\includegraphics[width=\columnwidth]{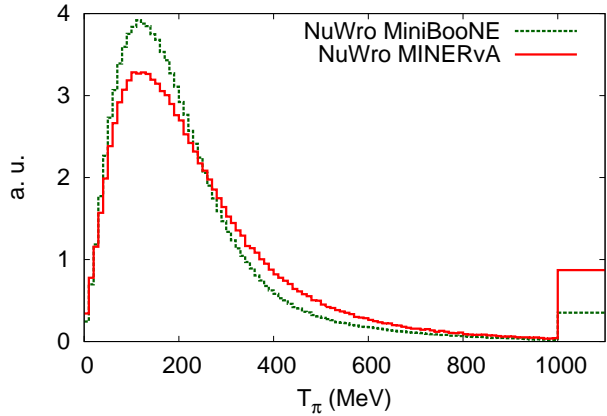}
\caption{(Color online) Spectrum of pion kinetic energy for MiniBooNE and MINERvA as predicted by NuWro without FSI effects.
The last bin combines pions with kinetic energies above 1~GeV.}\label{fig5}
\end{figure}

In order to establish the leading systematic errors we identified a
dominant dynamical process giving rise to the signal in both experiments.
The contributions from the RES channel 
usually exceed 85-90\%,
see Fig. \ref{fig4}.
We conclude that most of the MiniBooNE and MINERvA signal events originate from the same physical 
processes. The pion kinetic energy distribution produced in RES process before 
FSI is quite similar for two experiments, see  Fig. \ref{fig5}.
Thus we expect that the impact of pion FSI effects is also similar in both cases.
In Ref. \cite{Eberly:2014mra} we found an information that according to GENIE $24\%$ of the MiniBooNE signal events correspond 
to $W>1.4$~GeV. In NuWro simulations the fraction is 23\%.

For the NuWro ratio results we used a simplified MC systematic error analysis based
on uncertainties in the RES process, which should cover the leading error of 
MC predictions.
Two error sources are taken into account:
\begin{enumerate}
\item $\Delta$ production rate uncertainty driven by $C_5^A$ and $M_{A\Delta}$ parameters.
\item $\Delta$ decay uncertainty coming from pion angular correlations.
\end{enumerate}

We varied the axial coupling of the $\Delta$ resonance $C_5^A(0)=1.19\pm 0.08$ and 
$M_{A\Delta}=0.94\pm 0.03\;$(GeV) within the limits found in Ref. \cite{Graczyk:2009qm} and
treat the maximum variation as a systematic errors $\delta_{C_5^A}$, 
$\delta_{M_{A\Delta}}$. We compared descriptions of the angular anisotropy reported by ANL and BNL experiments 
also took the maximum variation from both parameterizations as another
systematic error $\delta_{decay}$. 
We combined these errors in quadrature and obtained the estimate of the total error in NuWro simulations 
$\delta_{MC}=\sqrt{\delta_{C_5^A}^2+\delta_{M_{A\Delta}}^2+\delta_{decay}^2}$.

\begin{figure}[!htb]
\centering\includegraphics[width=\columnwidth]{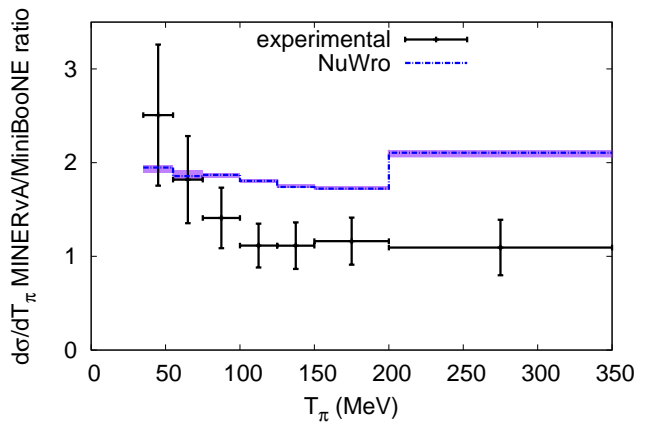}
\caption{(Color online) Ratios of differential cross sections in pion kinetic energy from MINERvA and MiniBooNE experiments
together with NuWro predictions.
}\label{fig6}
\end{figure}

In Fig. \ref{fig6}  we show the final results for
\[ 
 \frac{(\frac{d\sigma}{dT_\pi})^{MINERvA}         }
      {(\frac{d\sigma}{dT_\pi})^{MiniBooNE}         }\ \ \
\]
where the experimental results are compared to the NuWro predictions. 
NuWro central results are
obtained with BNL angular correlations and default values of $C_5^A(0)=1.19$ and $M_{A\Delta}=0.94\;$~GeV.

It is essential to look independently for the shape of the ratio.
Differences in shape are perhaps more important than discrepancy in the overall scale 
which can be due to uncertainties in overall cross section normalizations in two experiments.
In the case of the 
MINERvA measurement the overall normalization error can be estimated to be $\sim 15\%$ (flux error, correction for muon angles exceeding $20^o$,
detector effects)\  {\footnote{Brandon Eberly, private communication.}}. MiniBooNE normalization error should be similar in size.

\begin{figure}[!htb]
\centering\includegraphics[width=\columnwidth]{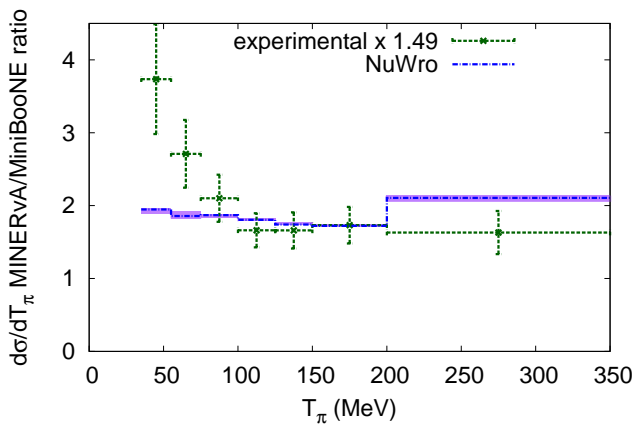}
\caption{(Color online) The same as in Fig. \ref{fig6} but with experimental results rescaled by a factor of $\eta = 1.49$.
}\label{fig7}
\end{figure}

In order to compare the shapes of two ratios we introduce a scaling factor $\eta$ and
found its value by trying to adjust the experimental results and NuWro predictions. 
%We took into account all the
%kinetic energy bins and the first 8 bins in the angular distribution, see the discussion in Sec. \ref{sec:data}.
We obtained the best-fit value $\eta=1.49\pm 0.15$. The value of $\eta$ is surprisingly large
compared to the estimated normalization error from two experiments. Also the shapes of measured and calculated with NuWro ratio as a function of pion kinetic energy are different, see Fig. \ref{fig7}, where
rescaled experimental result together with NuWro predictions 
are shown.

%The value of $\chi^2_{min}/D.O.F.$ for the first fit pion kinetic energy ratio suggests there is no statistically significant 
%shape discrepancy between
%Monte Carlo and experimental data even if at first sight it seems to be so. 

In the past GENIE \cite{Andreopoulos:2009rq} and
GiBUU \cite{Lalakulich:2012cj} also had problems with understanding $\pi^+$ production data. In the case of GiBUU
in \cite{Lalakulich:2012cj} a paradoxical conclusion was drawn that the MiniBooNE data is reproduced better if FSI effects are neglected.
There are differences in the underlying SPP and FSI models in all the generators and the model that is implemented in 
NuWro can be improved in many respects.
Even though, it seems unlikely that such a large data/Monte Carlo discrepancy is caused only by the deficiencies of the 
NuWro treatment of neutrino pion production.
The main argument is that in the cross section ratios all the implemented models defects should roughly cancel each other because in both cases 
the dominant dynamical mechanism: $\Delta$ excitation and decay is exactly the same and also the FSI effects are expected to be very similar.

\subsection{Pion angular distribution}

We studied also pion angular distribution from both experiments. 
In the case of the MiniBooNE experiment pion angular
distribution the points are taken from M. Wilking PhD thesis \cite{Wilking:2009zza}.
This data must be considered with precaution, as even if they are public they are not official MiniBooNE result.

The first problem is that the MiniBooNE detector has little sensitivity to the pion direction near the Cherenkov threshold 
at $T_\pi\sim$70 MeV. This results in a cutoff at pion kinetic energy of 150 MeV in the double-differential cross section 
presented in Tab. XVII of Ref. \cite{AguilarArevalo:2010bm}.
The second problem is that Ref. \cite{Wilking:2009zza} does not include improvements coming from better algorithms to separate muons and charged pions which have some impact on unfolded pion differential cross section.

For the second problem we looked at overlapping kinematical region from Ref. \cite{Wilking:2009zza} and the MiniBooNE paper \cite{AguilarArevalo:2010bm} 
for pion double differential cross section results. The agreement is on the level of  $\sim 1-6$\%.

As for the first problem we used NuWro Monte Carlo generator to estimate the range of pion production angles $\Theta_\pi$, 
for which $T_\pi<$150 MeV events do not dominate. 
We noticed a general pattern that more energetic pions preferably move in the forward and less energetic in the backward hemispheres. This can be 
understood as a relativistic effect on a mostly uniform distribution of pions in the $\Delta$ rest frame boosted to the laboratory frame.
%\begin{figure}[!htb]
%\centering\includegraphics[width=\columnwidth]{plots/fig3}
%\caption{(Color online) Contributions to MiniBooNE differential cross section in pion production angle 
%with signal fractions coming from low energy pions as estimated by NuWro. MiniBooNE data points are taken from Ref. \cite{Wilking:2009zza}. On the left from  vertical line at $\Theta_\pi = 70^o$ the contribution
%from low kinetic energy pions is small.}\label{fig3}
%\end{figure}
%The fraction of events with the low pion kinetic energy in the function of $\Theta_\pi$ is shown Fig. \ref{fig3}. 
We observed that for the pion production angles $\lesssim$70$^o$ the fraction 
of near-threshold
pions does not exceed 13\% and the contribution of pions with kinetic energies below 150 MeV is smaller than $\sim 50\%$. It is plausible that the data from Ref. \cite{Wilking:2009zza} 
is trustworthy for $\Theta_\pi \lesssim 70^o$ and we compared them with the MINERvA results. As before, we calculated ratios
of experimental results and Monte Carlo prediction. The conclusion is that there is a significant
disagreement in shape. NuWro predicts that the ratio should be roughly equal 2 for $\Theta_\pi\in [0^o,70^0]$. On the contrary, the experimentally measured ratio $(\frac{d\sigma}{dT_\pi})^{MINERvA}        / (\frac{d\sigma}{dT_\pi})^{MiniBooNE} $     
shows a strong drop from the value $\sim 2.5$ for $\Theta_\pi=0^o$ to $\sim 0.8$ for  $\Theta_\pi\approx 50^o$.

\section{Conclusions}
\label{sec:conclusions}

A comparison of experimental $\pi^+$ production data from MiniBooNE and MINERvA experiments reveals that there is a large (a factor of $1.49$) normalization
discrepancy between two measurements. There are also noticeable differences in the measured shapes of differential cross section
in pion kinetic energy.  Unfortunately, the MiniBooNE Cherenkov detector
does not provide us with reliable angular distribution due to the near-threshold effects and we can not give any definite conclusions for this observable.

We are still far away from a good understanding of SPP channels in the neutrino scattering in the $\Delta$ region. 
Interpretation of the old ANL and BNL
deuteron target experiments is not straightforward  because of apparent differences in measured cross sections (see, however a 
discussion in Refs. \cite{Graczyk:2009qm} and \cite{Wilkinson:2014yfa}) and problems with modeling nonresonant contribution \cite{Graczyk:2014dpa}.
It is clear that more dedicated experimental effort aiming to measure the pion production reactions
together with nuclear effects is needed.

\subsection*{Acknowledgements}
%We thank M. Wilking for explaining us the problems with reconstruction of low energy pions in the MiniBooNE experiments.
We thank B. Eberly and S. Dytman for useful comments on the MINERvA  and M. Wilking on the MiniBooNE experiments. 
We also thank Barbara Szczerbinska for a warm hospitality at the CETUP*2014 (Center for Theoretical Underground Physics and Related Areas) 
workshop in South Dakota where the idea of this study was conceived. Both the authors were supported by NCN grant UMO-2011/M/ST2/02578.

\bibliographystyle{aip}

\bibliography{bibdrat}
\end{document}